\documentclass[journal]{IEEEtran}
\usepackage{threeparttable}
\usepackage{enumerate} 
\usepackage{multirow}
\usepackage{color}
\usepackage{amssymb}
\usepackage{cite}

\usepackage{listings}
\usepackage{xcolor}
\usepackage{pifont}

\ifCLASSINFOpdf
  \usepackage[pdftex]{graphicx}
  \graphicspath{./}
 \DeclareGraphicsExtensions{.pdf,.jpeg,.png}
\else
\fi

\usepackage{amsmath}
\usepackage{bm}
\usepackage{algorithm}
\usepackage{algorithmicx}
\usepackage{algpseudocode}

\usepackage{array}

\ifCLASSOPTIONcompsoc
 \usepackage[caption=false,font=normalsize,labelfont=sf,textfont=sf]{subfig}
\else
 \usepackage[caption=false,font=footnotesize]{subfig}
\fi

\usepackage{url}

\begin{document}
%
\title{DNA-HHE: \underline{D}ual-mode \underline{N}ear-network \underline{A}ccelerator for \underline{H}ybrid \underline{H}omomorphic \underline{E}ncryption on the Edge}
%
%
%

\author{\IEEEauthorblockN{
Yifan Zhao${}^{*}$,
Xinglong Yu${}^{*}$,
Yi Sun${}^{*}$,
Honglin Kuang${}^{*}$,
Jun Han${}^{*}$}\\
\IEEEauthorblockA{
\textit{${}^{*}$State Key Laboratory of Integrated Chips and Systems, Fudan University} \\
Shanghai, China}
}

%
%

\markboth{}%
{Shell \MakeLowercase{ZHAO \textit{et al.}}: ARV-Q: An Adaptive RISC-V Vector Processor for Unified Post-Quantum Cryptography Support and Side-Channel Protection}
%



\maketitle

\begin{abstract}
Fully homomorphic encryption (FHE) schemes like RNS-CKKS enable privacy-preserving outsourced computation (PPOC) but suffer from high computational latency and ciphertext expansion, especially on the resource-constrained edge side. 
Hybrid Homomorphic Encryption (HHE) mitigates these issues on the edge side by replacing HE with lightweight symmetric encryption for plaintext encryption, such as the Rubato cipher for the HHE variant of RNS-CKKS, yet it introduces transciphering overhead on the cloud.
The respective strengths and limitations of FHE and HHE call for a dual-mode HHE solution with flexible algorithm switching ability.
This paper presents DNA-HHE, the first dual-mode HHE accelerator with near-network coupling for edge devices.
DNA-HHE supports both edge-side RNS-CKKS and Rubato within a unified architecture driven by flexible custom instructions.
To realize a compact implementation for the edge side, we propose a DSP-efficient modular reduction design, a compact multi-field-adaptive butterfly unit, and parallel scheduling schemes of Rubato with a high degree of resource sharing.
DNA-HHE is designed with network protocol packaging and transmission capacities and directly coupled to the network interface controller, achieving reduced overall latency of edge-side PPOC by 1.09$\times$ to 1.56$\times$.
Our evaluations on the ASIC and FPGA platforms demonstrate that DNA-HHE outperforms the state-of-the-art single-mode designs in both edge-side RNS-CKKS and symmetric cipher with better computation latency and area efficiency, while offering dual-mode functionality.
\end{abstract}


%
\IEEEpeerreviewmaketitle

\section{Introduction}
\begin{figure}[t!]
  \centering
  \includegraphics[width=1.0\linewidth]{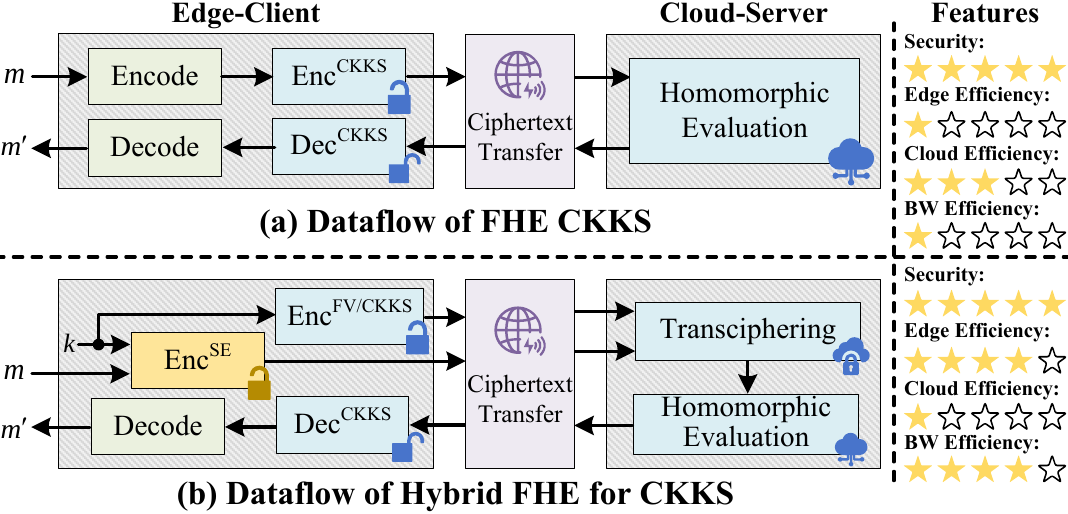}
  \caption{End-to-end workflow of traditional FHE and Hybrid HE scheme in PPOC with edge computing paradigm.}
  \label{fig:FHEvsHHE}
  \vspace{-10pt}
\end{figure}

Fully homomorphic encryption (FHE), enabling computations on ciphertexts without decryption, has emerged as a cornerstone for privacy-preserving outsourced computation (PPOC).
Current mainstream FHE algorithms include BFV \cite{FanV12} and BGV\cite{brakerski2014leveled} for integer arithmetic, CKKS \cite{CheonKKS17} for approximate arithmetic, and TFHE \cite{Joye22} for boolean circuits.
Among FHE schemes, the Residue Number System (RNS) variant of CKKS (RNS-CKKS) \cite{CheonHKKS18} has gained particular prominence as its approximate arithmetic support makes it ideal for real/complex-number operations such as those in signal processing and machine learning.
The workflow of RNS-CKKS PPOC is shown in Fig. \ref{fig:FHEvsHHE} (a), which mainly involves the homomorphic Encryption/Decryption (Enc/Dec) on the edge, the homomorphic evaluation on the cloud, and the ciphertext transfer between edge and cloud through networking equipment like network interface controller (NIC).
Despite the promising potential of FHE-based PPOC, FHE schemes like RNS-CKKS face two major challenges at resource-constrained edge devices, as shown in the right of Fig. \ref{fig:FHEvsHHE} (a).
First, HE deployment suffers from significant computational overhead and latency due to the complicated HE operations, particularly the HE Enc over all RNS domains.
Second, the noise introduction mechanism of HE intrinsically causes serious ciphertext expansion, with sizes typically hundreds of times larger than plaintext, and this issue becomes even more severe with short-length messages.
As a result, edge devices face massive bandwidth (BW) pressure for ciphertext transfer both in internal interconnections and externally to the cloud, leading to low transfer efficiency.

The drawbacks of FHE in slow Enc and ciphertext expansion on the edge side have spurred the development of Hybrid Homomorphic Encryption (HHE) schemes \cite{Dobraunig0HRSW23, HaKLLS22, CosseronHMS22}.
The key idea of HHE is to combine HE with lightweight symmetric encryption (SE) to create a more efficient scheme suitable for edge-side practical applications, such as the Rubato SE \cite{HaKLLS22} tailored for RNS-CKKS and the Pasta SE \cite{Dobraunig0HRSW23} tailored for BFV/BGV.
Fig. \ref{fig:FHEvsHHE} (b) demonstrates the HHE variant of RNS-CKKS-based PPOC.
On the edge, HHE utilizes the SE Enc to encrypt all the plaintexts in segments instead of the expensive HE Enc, significantly reducing computational overhead and latency.
Moreover, since the SE cipher does not cause ciphertext expansion, the ciphertext size is considerably reduced, leading to substantially lower BW requirements and decreased ciphertext transfer latency.
HHE mitigates the limitations of FHE on the edge but introduces cloud-side drawbacks.
The cloud must convert the SE ciphertext into the homomorphic ciphertext, namely transciphering, which adds significant computational overhead.

After analyzing both FHE and HHE schemes, it becomes evident that their respective strengths and limitations call for an adaptable approach for edge-side PPOC deployment.
This necessitates the demand for a \textbf{dual-mode HHE solution}, which can dynamically select the optimal edge-side encryption scheme based on current conditions: 
when edge devices face latency or bandwidth constraints, selecting HHE with SE Enc reduces computational and bandwidth demands;
when the cloud experiences surging requests and elevated latency, switching to pure HE eliminates the cloud transciphering overhead.
However, challenges exist in deploying dual-mode HHE on the edge side.
\textbf{(1) Lacking of existing works supporting dual-mode HHE.}
Existing hardware implementations for edge-side PPOC exclusively concentrate on the single-mode acceleration of pure HE schemes such as \cite{WangYHZMSL25, KriegerHMR24, AzadYAPTJ23}, or a single symmetric cipher such as \cite{cryptoeprint:2024/1919}, leaving the dual-mode combination unexplored.
\textbf{(2) Increased area overhead due to dual-mode functionality.}
Given the distinctions between pure HE and SE cipher in terms of numerical fields and computation patterns, dual-mode HHE may result in increased area overhead, hindering its deployment on resource-constrained edge devices.
\textbf{(3) Large transmission overhead of expanded ciphertext.}
Our evaluation shows ciphertext transmission between the accelerator and NIC accounts for 16.7\% to 50.6\% of the latency overhead in edge-side end-to-end PPOC under different interconnect bus specifications.
However, existing designs are implemented as standalone accelerators and overlook the optimization for expanded ciphertext transfer inherent between accelerators and NIC, and this standalone approach leads to diminished efficiency.

In response to the identified challenges,  we present DNA-HHE, a dual-mode HHE accelerator with near-network coupling for edge devices.
DNA-HHE supports both the edge-side RNS-CKKS \cite{CheonHKKS18} and the latest Rubato SE cipher \cite{HaKLLS22} tailored for the corresponding HHE variant, within a unified compact architecture that is tightly coupled to the NIC.
Our contributions are as follows:
\begin{itemize}
\item 
\textbf{Programmable Dual-mode HHE.}
We propose the \textbf{\textit{first}} dual-mode HHE accelerator for edge-side RNS-CKKS and Rubato SE cipher in a unified architecture.
A set of task-level instructions is introduced to flexibly drive the data flow with out-of-order execution to reduce latency.
\item 
\textbf{Compact Architecture.}
For resource-limited edge devices, we emphasize enhancing hardware efficiency through the proposed DSP-efficient modular reduction method, compact multi-field-adaptive butterfly unit (BFU), and parallel scheduling schemes of Rubato with a high degree of resource sharing.
\item 
\textbf{Near-network Acceleration.}
To reduce the ciphertext transfer overhead on the edge side, DNA-HHE is designed as a tightly coupled accelerator to the NIC, with instruction-driven network protocol packaging and transmission capabilities.
Compared to the standalone solutions, our near-network approach achieves $1.09\times$ to $1.56\times$ speedup in overall latency of edge-side end-to-end PPOC under different interconnect bus specifications.
\item
\textbf{Evaluation.}
We implement DNA-HHE respectively in ASIC and FPGA platforms.
Experimental results show that DNA-HHE outperforms the state-of-the-art single-mode counterparts in both RNS-CKKS and symmetric cipher with better computation latency and area efficiency, while offering dual-mode functionality.
\item
\textbf{Insights of RNS-CKKS vs. Rubato.}
The respective suitable scenarios of RNS-CKKS and Rubato are identified based on detailed performance comparison and analysis.
\end{itemize}

\section{Background and Motivation}
\subsection{Notation}
We denote the ring of integers modulo a prime $q$ by $\mathbb{Z}_q$.
Extending this, the ring of polynomials with integer coefficients modulo $q$ is symbolized as $\mathbb{Z}_q[X]$. 
Polynomials that are reduced modulo both $q$ and $2N$-th cyclotomic polynomial $X^N+1$ form the ring $R_{q}=\mathbb{Z}_q[X] /\left(X^N+1\right)$.

\subsection{Dual-mode HHE Overview}
Our target is the dual-mode HHE on the edge side, which is the combination of the RNS-CKKS and Rubato SE.
\subsubsection{Edge-side RNS-CKKS scheme}
\begin{figure}[htb]
	\centering
        \subfloat[m-to-ct: Encode+Encryption]{\label{fig:encryption}\includegraphics[width=0.7\linewidth]{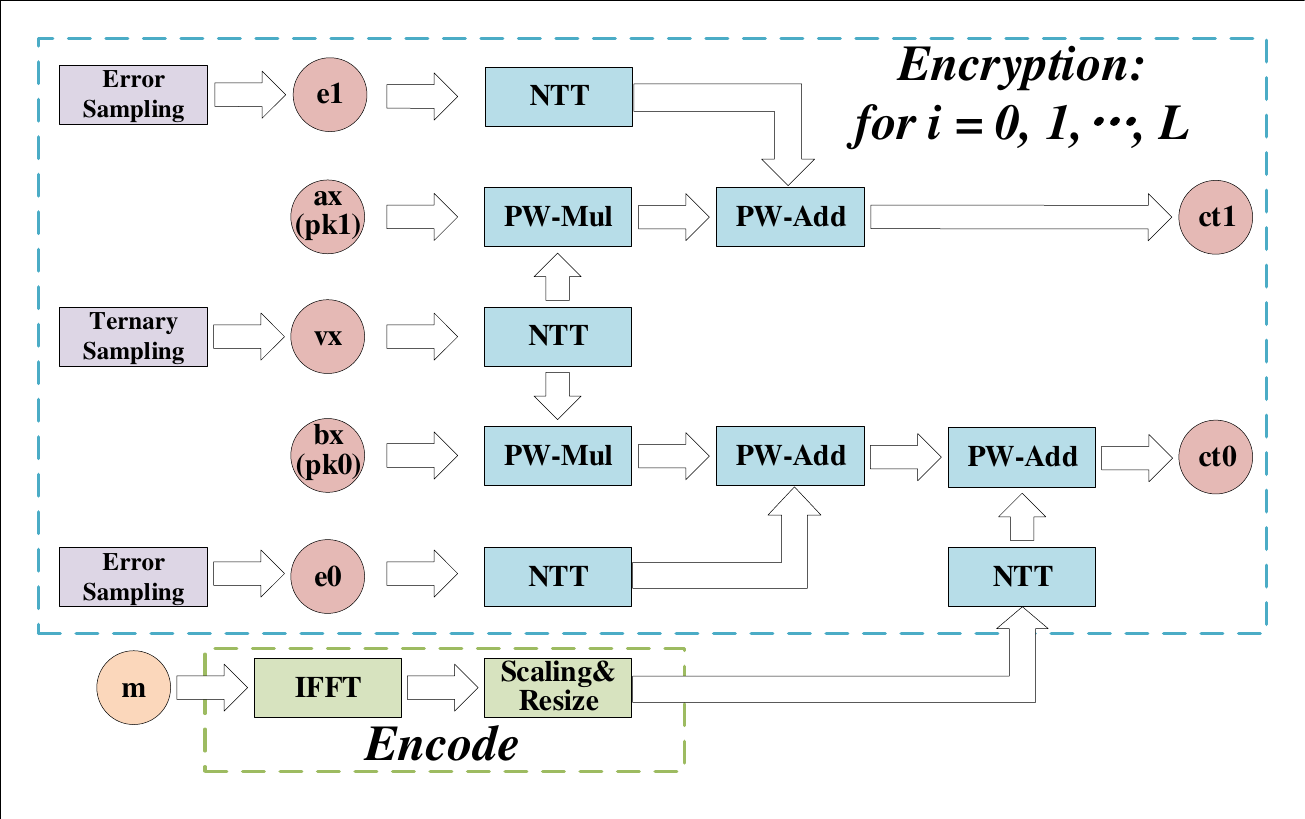}}
        \\
        \vspace{-10pt}
        \subfloat[ct-to-m: Decryption+Decode]{\label{fig:decryption}\includegraphics[width=0.7\linewidth]{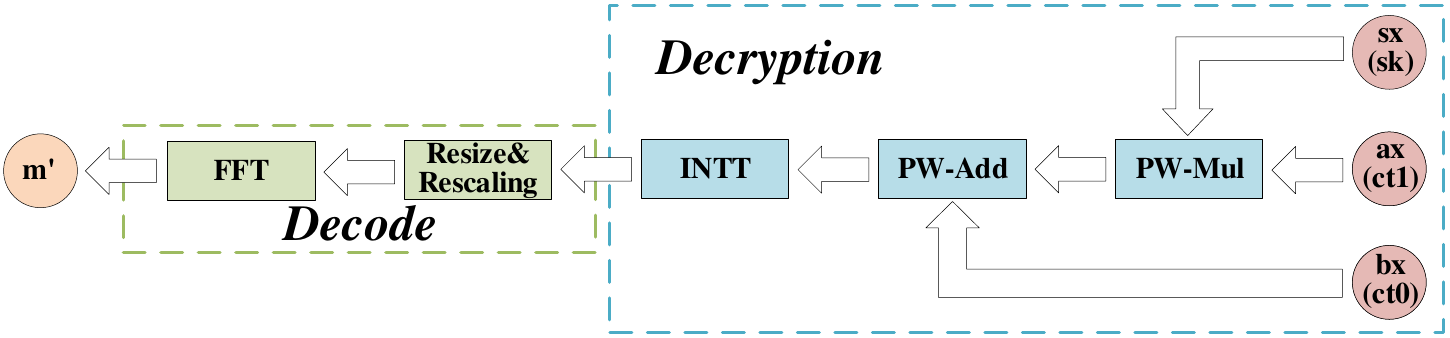}}
 \caption{Dataflow of edge-side RNS-CKKS scheme.}
 \label{fig:rns-ckks}
\end{figure}

As a variant of the CKKS \cite{CheonKKS17}, the RNS-CKKS \cite{CheonHKKS18} integrates the Residue Number System (RNS) on its foundation to enhance computational efficiency.
For a level $L$, the original CKKS employs the ciphertext modulus $Q_{L} = q^{L}$ as a power of the fixed base $q$.
In order to apply RNS, RNS-CKKS adopts a different ciphertext modulus $Q_{L}=\prod_{i=0}^{L} q_i$ with a set of pairwise co-prime modulus $\mathcal{B}=\left\{q_0, \ldots, q_L\right\}$ as the RNS basis.
The RNS basis is the approximate values of the original fixed base $q$ with error bounded by $q / q_{\ell} \in\left(1-2^{-\eta}, 1+2^{-\eta}\right)$ for $\ell=1, \ldots, L$, where $\eta$ is the bit precision and should be larger than the bit precision of plaintexts.

The dataflow of edge-side RNS-CKKS is shown in Fig. \ref{fig:rns-ckks}, which includes the message-to-ciphertext (m-to-ct) process, and the reverse ciphertext-to-message (ct-to-m) process.
In the Encode phase, a message $m \in \mathbb{C}^{L_{m}}$ with $L_{m} \leq N/2$ is converted into a plaintext polynomial $pt \in R_{q_{i}}$
by packing $L_{m} \times$ complex numbers represented in float point or fixed point to a single polynomial via inverse Fast Fourier Transform (IFFT),  scaling the result to the finite field $\mathbb{Z}_{q_{i}}$ and resizing to the plaintext polynomial $pt^{\left(i\right)} \in R_{q_{i}}$.
Decode reverses the Encode process.
The Encryption converts the plaintext polynomial  $pt^{\left(i\right)} \in R_{q_{i}}$ into the ciphertext polynomial tuple $\left(ct^{\left(i\right)}_0, ct^{\left(i\right)}_1\right) \in R^{2}_{q_i}$, and this process is performed on each RNS basis.
To reduce the computational complexity of the polynomial multiplication, the number theoretic transform (NTT) \cite{PoppelmannG12} is utilized, which enables using point-wise (PW) operations such as multiplication (Mul) and addition (Add) to accomplish the remaining process.
The Decryption process converts the ciphertext polynomial tuple $\left(ct^{\left(0\right)}_0, ct^{\left(0\right)}_1\right) \in R^{2}_{q_0}$ into the plaintext polynomial  $pt^{\left(0\right)} \in R_{q_{0}}$.


\subsubsection{Rubato Symmetric Encryption}
\begin{table}[]
\centering
\caption{Parameters of Rubato Cipher}
\label{tab:rubato_param}
\begin{tabular}{l|c|c|c|c|c|c}
\hline
Parameter & $\lambda$ & $v$ & $n$ & $l$ & $\lceil\log t\rceil$ & $r$ \\ 
\hline
\hline
Par-128S & 128 & 4 & 16 & 12 & 26 & 5 \\ \hline
Par-128M & 128 & 6 & 36 & 32 & 25 & 3 \\ \hline
Par-128L & 128 & 8 & 64 & 60 & 25 & 2 \\ \hline
\end{tabular}
\end{table}

\begin{figure}[!t]
  \centering
  \includegraphics[width=0.60\linewidth]{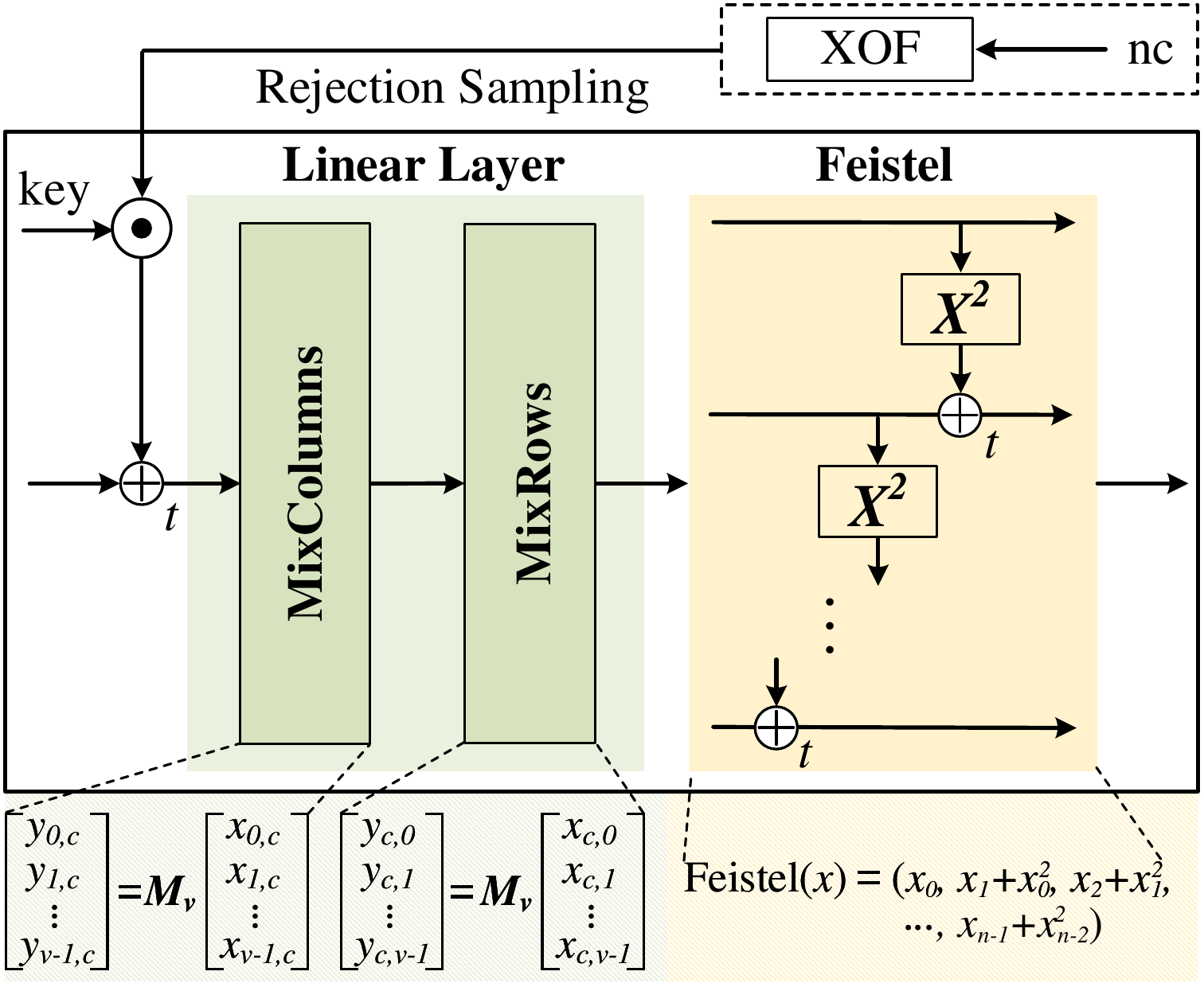}
  \caption{The round function of Rubato. Symbols $\bigodot$ and $\bigoplus$ respectively represent the PM-Mul and PW-Add over a finite field $\mathbb{Z}_t$}
  \label{fig:rubato}
  \vspace{-10pt}
\end{figure}

Rubato \cite{HaKLLS22} is the latest HE-friendly SE cipher designed for the HHE variant of RNS-CKKS \cite{CheonHKKS18}.
As shown in Table \ref{tab:rubato_param}, under the security level of $\lambda=128$ bit, it offers a flexible range of parameter options, including the block size $n=v^{2}$, the keystream output size $l$, the modulus $t$ and the round count $r$.
For $\lambda$-bit security, a symmetric key $\mathbf{k} \in \mathbb{Z}^{n}_{t}$ and a nonce $\mathrm{nc} \in \{0,1\}^{\lambda}$ are taken as input, and a keystream $\mathbf{k_{nc}} \in \mathbb{Z}^{l}_{t}$ is returned.
Rubato cipher includes calculations of $r$ rounds of the round function $\mathrm{RF}$, plus an additional final round function $\mathrm{Fin}$.
\begin{align}
\notag
\mathrm{RF[\mathbf{k}, nc},i] =&\mathrm{ARK[\mathbf{k}, nc},i] \circ \mathrm{MixColumns} \; \circ
\\
&\mathrm{MixRows} \circ \mathrm{Feistel}
\label{con:round_func}
\end{align}
The $i$-th round function $\mathrm{RF[\mathbf{k}, nc},i]$ is shown in Fig. \ref{fig:rubato} and defined in Eq. \ref{con:round_func}, in which the round key schedule is defined as $\mathrm{ARK[\mathbf{k}, nc},i](\mathbf{x}) = \mathbf{x} + \mathbf{k} \, \odot \,\mathbf{rc_{i}}$.
The $\mathbf{rc_{i}} \in \mathbb{Z}^{n}_{t}$ is obtained from rejection sampling where the random numbers are generated via the XOF function such as SHAKE \cite{dworkin2015sha} employing nonce $\mathbf{nc}$ as the random seed.
The final round function $\mathrm{Fin}$ replaces the Feistel operation in round function $\mathrm{RF}$ with a truncation function $\mathrm{Tr}_{n,l}$ which returns the first $\ell$ words of the $n$-word block.
Notably, all computational operations in Rubato, such as matrix-vector multiplication and PW-MUL/ADD, are performed within the finite field $\mathbb{Z}_{t}$.
 \vspace{-10pt}
\begin{figure}[htb]
	\centering
        \subfloat[Single-Mode Standalone FHE Accelerator]{\label{fig:old_acc}\includegraphics[width=0.95\linewidth]{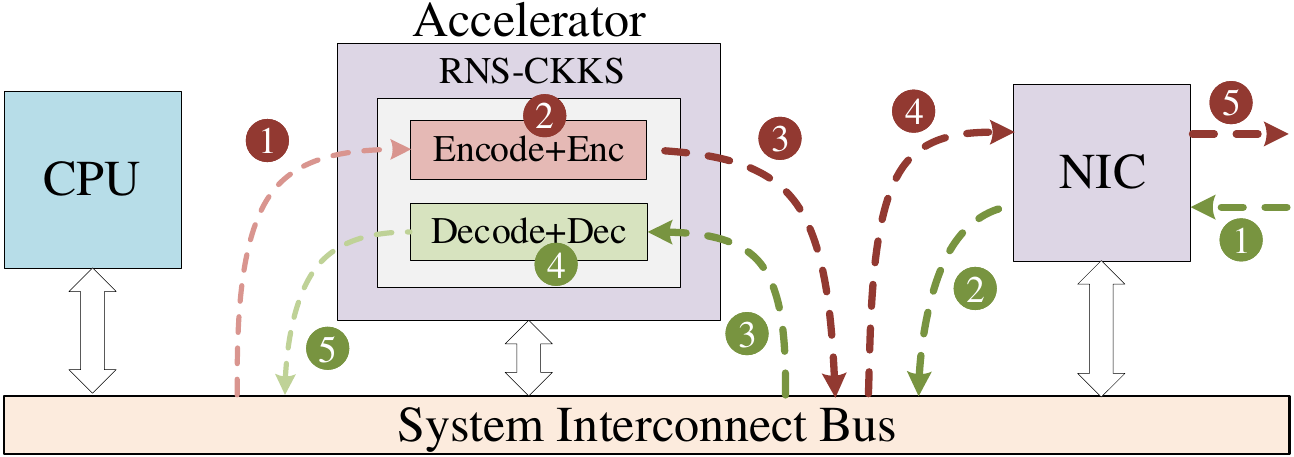}}
        \\
        \subfloat[Dual-mode Near-network HHE Accelerator]{\label{fig:new_acc}\includegraphics[width=0.95\linewidth]{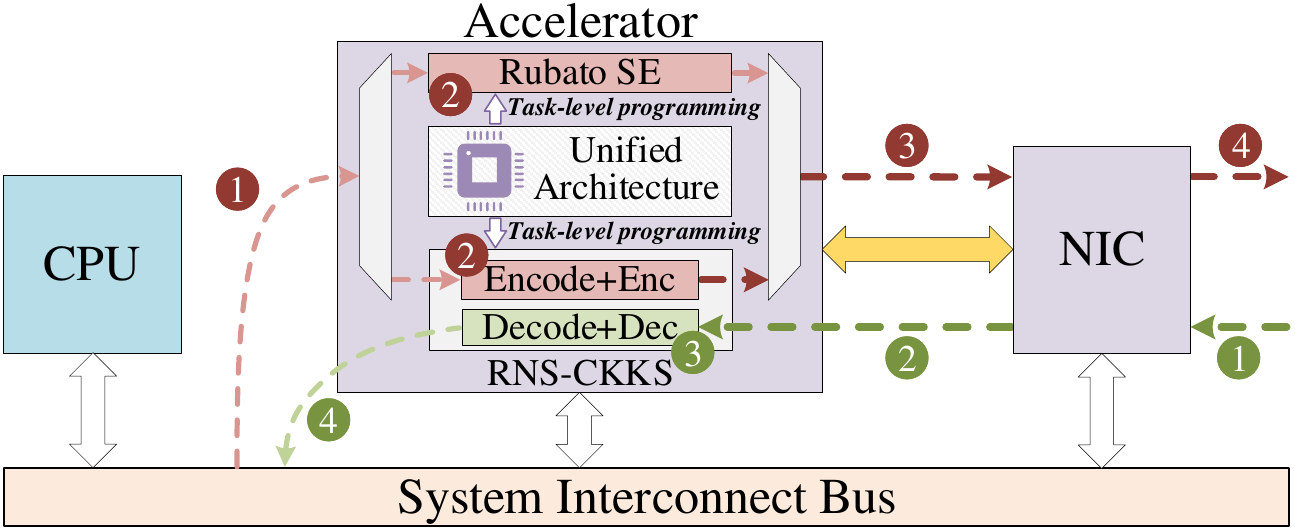}}
 \caption{High-level architecture of edge-side FHE accelerator and dataflow. The red path illustrates the steps involved in the m-to-ct conversion and network packets sending, and the green path shows the steps for network packets receiving and the ct-to-m conversion. Thicker lines with darker colors represent larger data transmission volumes.}
 \label{fig:acc_form}
 \vspace{-10pt}
\end{figure}

\subsection{Motivation}
There have been many recent works \cite{WangYHZMSL25, KriegerHMR24, AzadYAPTJ23} focusing on the edge-side acceleration of FHE, and all these works adhere to the design model of the single-mode standalone accelerator as depicted in Fig. \ref{fig:old_acc}.
Functionally, these accelerators are dedicated to HE operations at the edge without dual-mode support.
At the System-on-Chip (SoC) level, these designs operate in a standalone manner that focuses solely on the computation part of end-to-end PPOC applications and depends on the interconnect bus to communicate with other components.
However, in end-to-end PPOC, apart from the crypto computation, there exists significant expanded ciphertext transfer overhead between the accelerator and the NIC, as illustrated by \ding{174} \ding{175} in the red path and \ding{173} \ding{174} in the green path in Fig. \ref{fig:old_acc}, since ciphertext from the edge must be sent to the cloud for evaluation (red path), and the edge must receive results from the cloud for decryption (green path).
Additionally, since the transmission of network packets necessitates switching from user space to kernel space, this makes it challenging for ciphertext transfer to overlap with accelerator computation to hide transmission delays.
Consequently, this standalone design results in increased latency for end-to-end PPOC applications due to the extensive overhead of ciphertext transfer on the interconnect bus.

Alternatively, we propose a new design model, the dual-mode near-network HHE accelerator, for a more adaptable and faster edge-side PPOC application, as shown in Fig. \ref{fig:new_acc}.
Functionally, our design supports dual-mode HHE acceleration, allowing the user to dynamically select the most appropriate encryption scheme based on current conditions.
Additionally, given the limited resources at the edge, we prioritizes a unified and compact design through flexible task-level programming, algorithm-aware optimization, and extensive resource sharing.
To reduce the ciphertext transfer overhead, we adopt a near-network computing approach that the accelerator is directly coupled with the NIC and can handle packet en/de-capsulation and sending/receiving itself;
this enables the replacement of time-consuming ciphertext transfers that occur twice (Store+Load) via system bus with a single data transfer on a fast path, thus effectively reducing the overall latency of end-to-end PPOC applications.

\begin{figure*}[htb]
  \centering
  \includegraphics[width=0.8\linewidth]{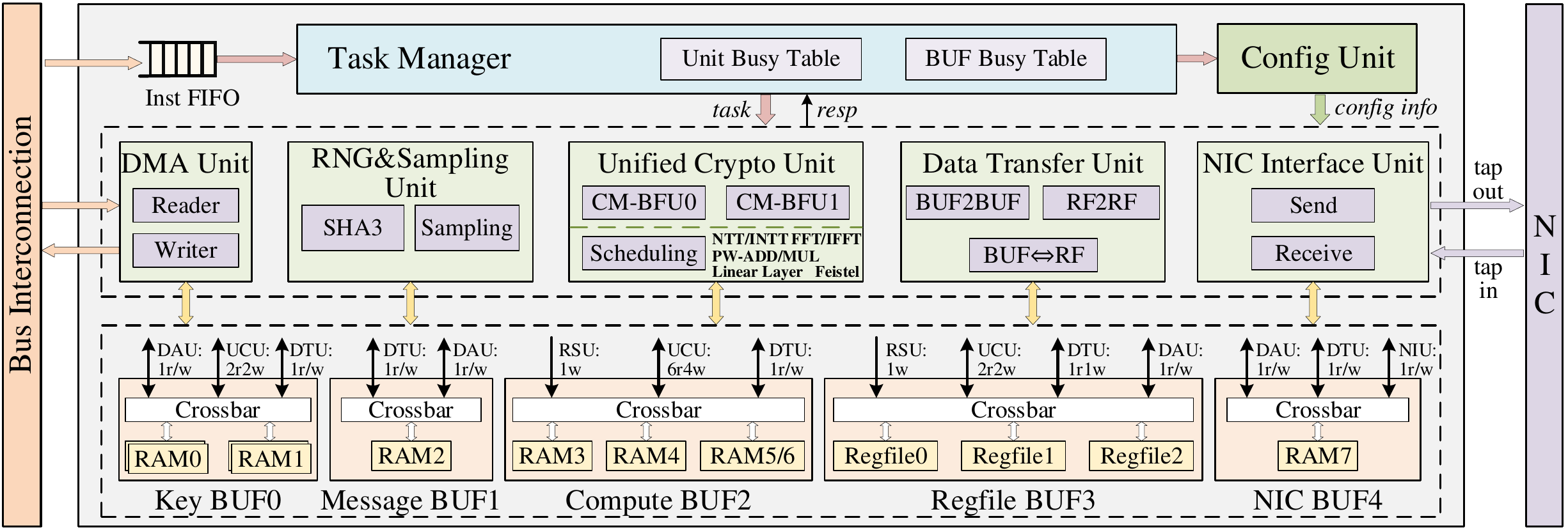}
  \caption{Overview of the high-level hardware architecture of DNA-HHE.}
  \label{fig:top_level}
\end{figure*}
\section{DNA-HHE Design}
\subsection{Architecture Overview}
Fig. \ref{fig:top_level} shows the architecture of the DNA-HHE accelerator.
DNA-HHE is driven by the proposed custom instruction set to enable flexible dual-mode switching, dynamic adjustment of algorithm parameters, and network packet transmission with NIC.
The custom instructions can be divided into two categories, the configuration type and the functionality type.
The former is used to configure the parameter registers within the Config Unit, such as the modulus $q_{i}$ and $t$ in the RNS-CKKS and Rubato, and the header size of network packets, while the latter is used to perform functional tasks.
Task Manager is designed to handle conflicts and manage task dispatching to the functional units, which include the DMA Unit (DAU) for system bus access, RNG\&Sampling Unit (RSU) for random number generation and sampling, Unified Crypto Unit (UCU) for crypto operations across multiple fields, Data Transfer Unit (DTU) for moving data between BUFs, and NIC Interface Unit (NIU) for network protocol packaging and transmission through 64-bit-width direct connection with NIC.
Five buffer clusters are introduced as the local memory.
RAM0/1/3/4 are composed of two dual-port banks.
RAM0/1 are used to store keys in the RNS-CKKS, while RAM3/4 are primarily used for FFT/IFFT and NTT/INTT operations in a ping-pong manner.
RAM5 and RAM6 are respectively used to store the twiddle factors in FFT/IFFT and NTT/INTT.
DNA-HHE supports up to 56-bit-width and 28-bit-width modular operations in RNS-CKKS and Rubato, respectively, and complex numbers based on 29-bit-width signed fixed point are employed for FFT/IFFT with bit precision $\eta$ set to 26.

\subsection{Compact Multi-field-adaptive BFU Design}
The BFU serves as the computing workhorse of FHE accelerators. 
In this work, the design of the BFU  targets two goals: compactness for edge-side deployment and multi-field adaptability for dual-mode HHE operations.
\subsubsection{DSP-efficient modular reduction}
For the most resource-demanding modular reduction in RNS-CKKS, we propose a DSP-efficient design.
The original $k$-bit Barret reduction \cite{barrett1986implementing} requires two $k$-bit-width multiplications: one for the precomputed value $\mu_{i}=\left\lfloor\frac{2^{2 k}}{q_{i}}\right\rfloor$ and another for the modulus $q_{i}$.
Since each modulus $q_{i}$ in our chosen RNS basis $\mathcal{B}$ follows the format in Eq. \ref{con:q_format}, multiplying by $q_{i}$ can be efficiently performed through bit-shifts, additions and multiplying by $bnd_{i}$.
Based on our evaluation, with $k=54$ and $\eta=26$, a 10-bit width for $bnd_{i}$ yields 49 primes satisfying the requirements for the RNS base $q_{i}$, and thus a 10-bit width is sufficient for practical application.
The multiplication with $\mu_{i}$ can be optimized using a similar method because we prove that $\mu_{i}$ follows the format in Eq. \ref{con:mu_format}.
\begin{equation}
    q_{i} = 2^{k} - 2N\times bnd_{i} + 1
    \label{con:q_format}
\end{equation}
\begin{equation}
    \mu_{i}=2^{k}+\Delta_{i}-1
    \label{con:mu_format}
\end{equation}
The proof that $\mu_{i}$ adheres to this format can be transformed into proving that a unique $\Delta_{i}$ exists, satisfying $\mu_{i}=2^{k}+\Delta_{i}-1=\left\lfloor\frac{2^{2 k}}{q_{i}}\right\rfloor$, and this verification can be further converted into proving that there is a unique $\mu_{i}$ such that $\mu_{i}=2^{k}+\Delta_{i}-1$ meeting the conditions shown in Eq. \ref{con:mu_constraint}.
\begin{equation}
    q_{i}\times \mu_{i} < 2^{2k} \;\&\&\; q_{i}\times (\mu_{i}+1)>2^{2k}
    \label{con:mu_constraint}
\end{equation}
Substituting Eq. \ref{con:q_format} and Eq. \ref{con:mu_format} into Eq. \ref{con:mu_constraint} leads to the derived constraints on $\Delta_{i}$ shown in Eq. \ref{con:delta_constraint}.
\begin{align}
\notag
\Delta_{i} &> \delta_{1}= \frac{(bnd_{i}\times 2N-1)\times 2^{k}}{2^{k}-bnd_{i}\times 2N+1}
\\
\Delta_{i} &< \delta_{2} = \frac{bnd_{i}\times2N\times(2^{k}-1)+1}{2^{k}-bnd_{i}\times 2N+1}
\label{con:delta_constraint}
\end{align}
\begin{figure}[htb]
  \centering
  \includegraphics[width=1.0\linewidth]{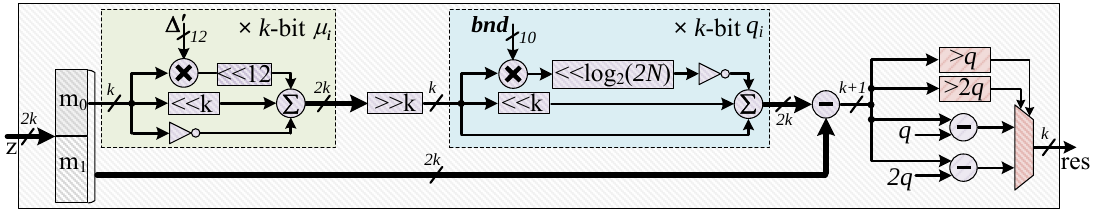}
  \caption{DSP-efficient Barret reduction architecture. 4-stage pipeline registers are omitted for simplicity.}
  \label{fig:reduction}
\end{figure}
Given that $\delta_{1}+1=\delta_{2}$ and $\Delta_{i}$ is an integer, we can conclude that there exists a unique $\Delta_{i}$ fulfilling Eq. \ref{con:delta_constraint};
therefore, we have proven the validity of Eq. \ref{con:mu_constraint} and Eq. \ref{con:mu_format}.
Moreover, with $k=54$ and 10-bit-width $bnd_{i}$, $\Delta_{i}$ can be represented as a 24-bit number in the format $\Delta_{i}$=\{$\Delta^{'}_{i}$, 12'b0\}.
Following the above optimization method, we propose a DSP-efficient Barret reduction unit as shown in Fig. \ref{fig:reduction}.
While maintaining support for scalable moduli up to 49 ones within the RNS $\mathcal{B}$, this architecture optimizes the originally required two $k$-bit multipliers into one $k$-bit by 12-bit multiplier and one $k$-bit by 10-bit multiplier, contributing to a 40\% reduction in DSP overhead for the CM-BFU in this work compared to \cite{WangYHZMSL25}.

\subsubsection{Multi-field adaptability}
\begin{figure}[t!]
  \centering
  \includegraphics[width=0.9\linewidth]{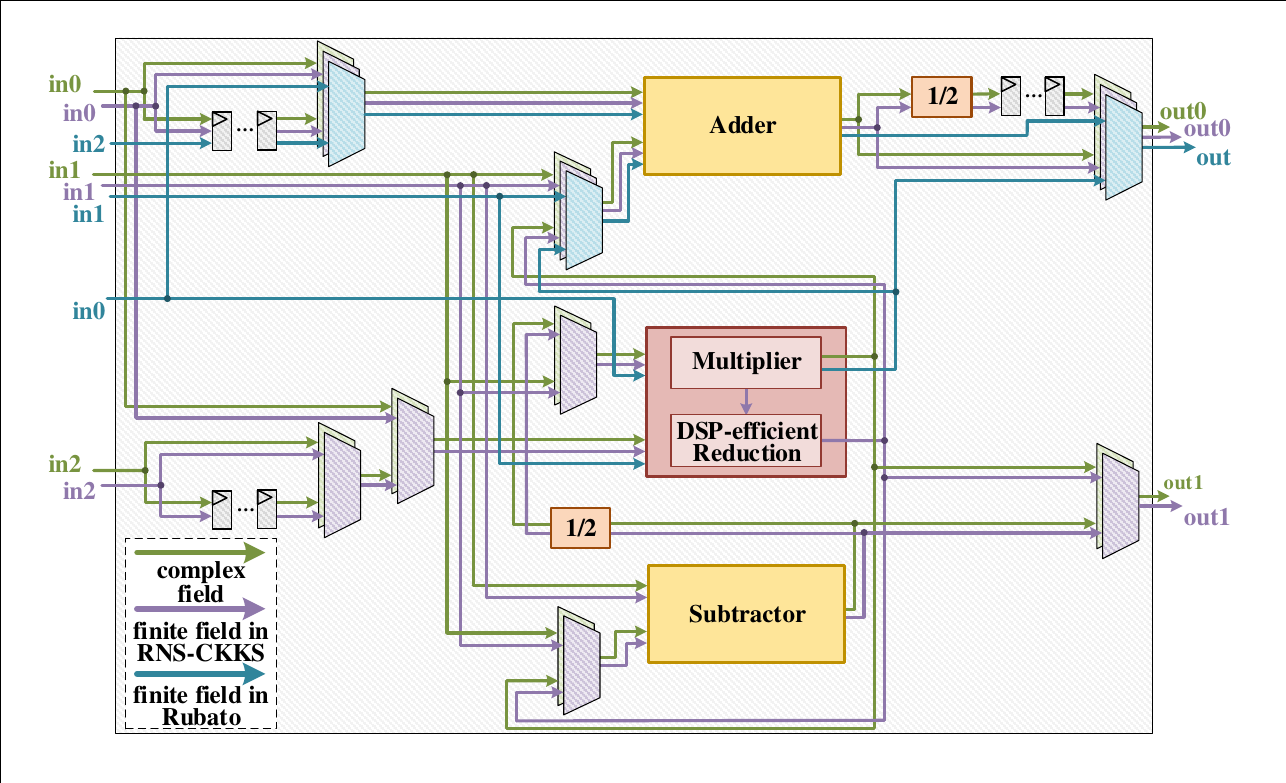}
  \caption{Compact Multi-field-adaptive BFU architecture.}
  \label{fig:BFU}
\end{figure}
\begin{figure}[t!]
  \centering
  \includegraphics[width=0.7\linewidth]{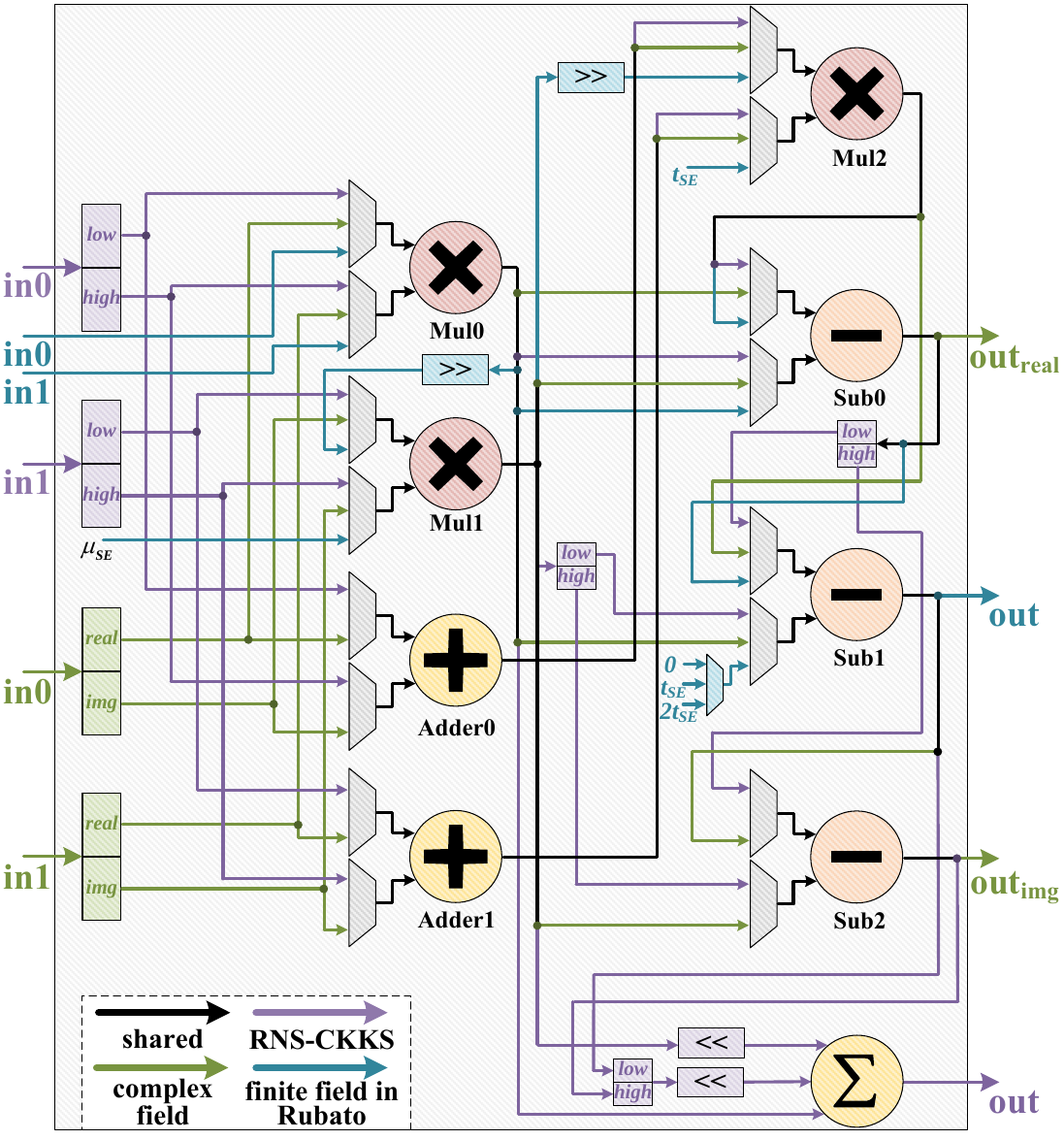}
  \caption{Multi-field-adaptive multiplier architecture. 3-stage pipeline registers are omitted for simplicity. \textit{\{high, low\}} respectively represent the higher half bits and lower half bits of a number. The subscripts \textit{real} and \textit{img} correspond to the real part and the imaginary part of a complex number, respectively.}
  \label{fig:Multiplier}
\end{figure}
To support operations in dual-mode HHE, the BFU is designed to perform the Cooley-Tukey/Gentleman-Sande butterfly, point-wise Mul/Add and multiply-accumulate (MAC) operations across multiple domains, including the fixed-point-based complex field and two distinct finite fields specific to RNS-CKKS and Rubato.
The architecture of the compact multi-field-adaptive BFU is shown in Fig. \ref{fig:BFU}, which consists of a DSP-efficient modular reduction units and multi-field-adaptive components including multiplier, adder, subtractor, and 1/2 units.
Fig. \ref{fig:Multiplier} illustrates the architecture of the multi-field-adaptive multiplier, which performs the integer multiplication in RNS-CKKS, the fixed-point multiplication in complex field, and the Barret modular multiplication in Rubato.
Karatsuba algorithm is adopted to optimize the multiplier usage of the integer multiplication in RNS-CKKS and the multiplication in complex field, reducing the need of 29-bit signed multipliers to 3, aligning with the requirement for Barret multiplication in the finite field of Rubato.
Other components follow the same design strategy as the multiplier, achieving multi-field adaptability while maintaining compactness by reconfigurable data path and fine-grained resource sharing.

\subsection{Operation scheduling of Rubato}
\begin{figure}[t!]
  \centering
  \includegraphics[width=1.0\linewidth]{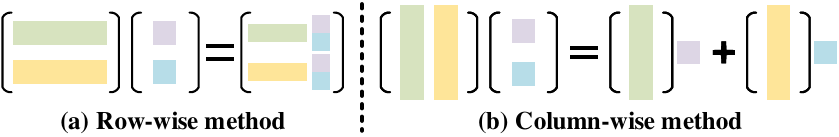}
  \caption{Approaches of matrix-vector multiplication.}
  \label{fig:MatrixMul}
\end{figure}
\begin{figure}[t!]
  \centering
  \includegraphics[width=0.7\linewidth]{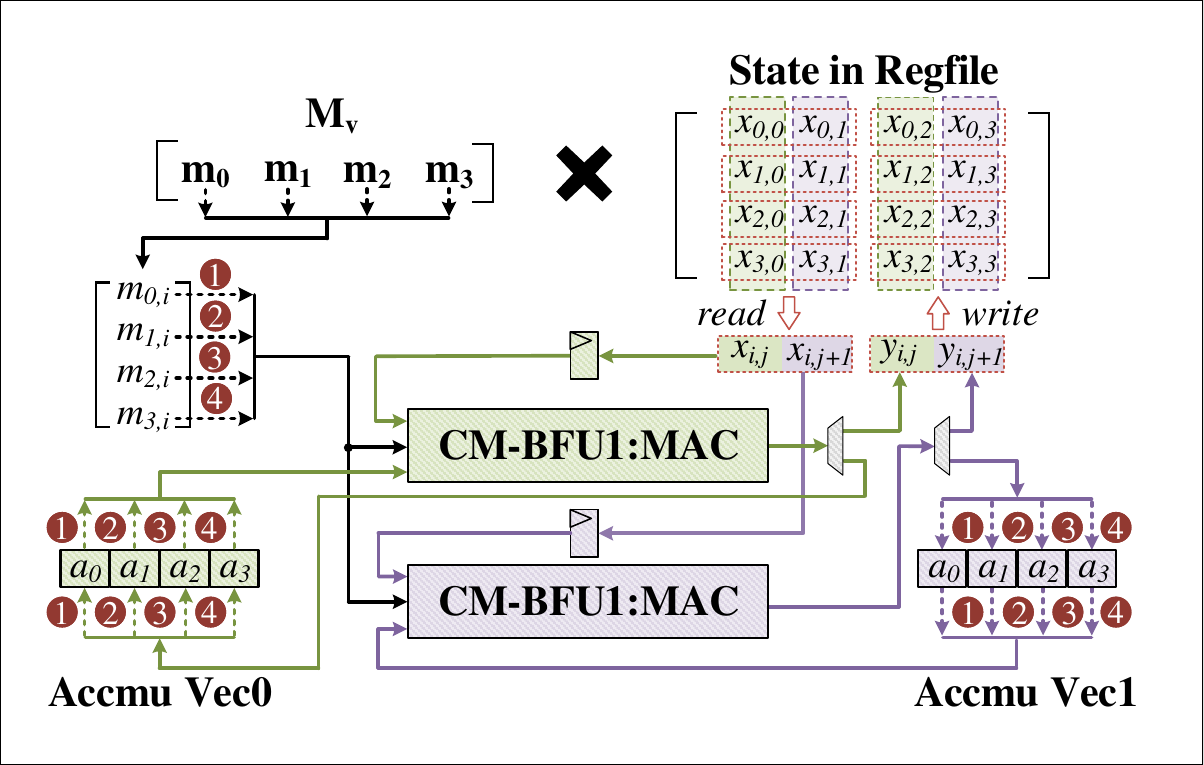}
  \caption{Column-oriented parallel scheduling of the MixColumns operation with $v=4$.}
  \label{fig:MixColumns}
\end{figure}
The Rubato SE involves relatively intricate Linear layer and Feistel operations.
To increase computing parallelism while fully reusing the underlying hardware, thus improving system compactness, efficient scheduling schemes for these operations are required.
\subsubsection{Linear layer}
The Linear layer is composed of the MixColumns and MixRows, which conduct the matrix-vector multiplication between the $v\times v$ matrix $\mathbf{M_v}$ and each column/row of the $v\times v$ state.
Being a constant circulant matrix, $\mathbf{M_v}$ can be entirely derived by keeping only its first column vector, $\mathbf{m0}$, in the Unified Crypto Unit.
The state is stored in the regifiles of BUF3, with each register element, shown by the red dashed box in Fig. \ref{fig:MixColumns}, holding two state coefficients to match the two-way parallelism of the CM-BFUs.

For the matrix-vector multiplications in the Linear layer, we propose a column-oriented parallel scheduling scheme.
Firstly, to maximize the usage of local column vector $\mathbf{m0}$ and reduce the access to regfiles in BUF3, we employ the column-wise method to perform the matrix-vector multiplication, as shown in Fig. \ref{fig:MatrixMul}b, instead of the row-wise method; in this way, the state coefficients in BUF3 only need to be read and written once during the Mix-Columns/Rows, thereby simplifying the access logic overhead of BUF3.
Besides, to ensure fully parallel computation of the two CM-BFUs, each CM-BFU individually performs the Mix-Column/Row operation on a separate column/row vector in the state.
To illustrate, the scheduling process of MixColumns under $v=4$ is shown in Fig. \ref{fig:MixColumns}.
Each extracted state coefficient is sequentially multiplied by the $v\times$ coefficients in the corresponding column vector from $\mathbf{M_v}$, with the product added to and result written to the corresponding element in the accumulation vector; this operation is efficiently executed through the $\mathbb{Z}_{t}$-field MAC computation of the CM-BFU.
When performing the MAC on the last coefficient in the state column vector, the current two MixColumn operations transition to the write-back phase, where the outputs of the two BFUs are concatenated and directly written into BUF3.
Through the column-oriented parallel scheduling, high computational efficiency of the Linear layer can be achieved while maintaining low hardware overhead.

\subsubsection{Feistel}
\begin{figure}[t!]
  \centering
  \includegraphics[width=0.6\linewidth]{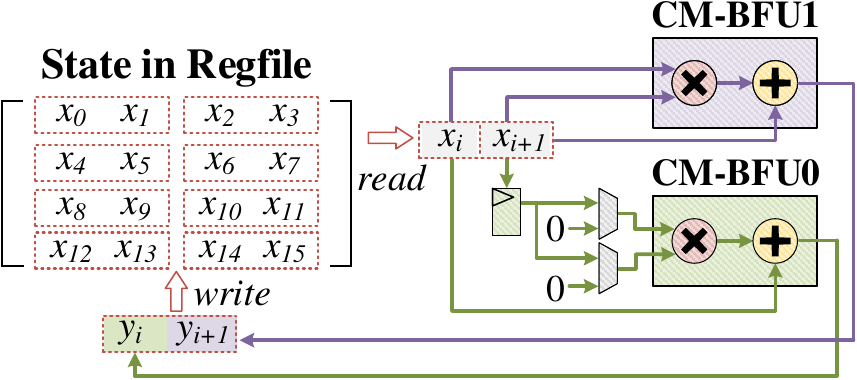}
  \caption{Pipelined-MAC scheduling of the Feistel operation with $v=4$, $n=16$.}
  \label{fig:Feistel}

\end{figure}
Feistel consists of $n-1$ sets of $(x^{2}_{i-1}+x_{i})$ operations.
As shown in Fig. \ref{fig:Feistel}, we propose a pipelined-MAC parallel scheduling for Feistel.
During each cycle, two coefficients $(x_{i}, x_{i+1})$ are retrieved, and the second one $ x_{i+1}$ is held in a register;
in this way, we can fully utilize the two CM-BFUs and perform two parallel MAC operations of $(x^{2}_{i-1}+x_{i})$ and $(x^{2}_{i}+x_{i+1})$ per cycle.

\vspace{-10pt}
\subsection{Workflow}
\begin{figure*}[htb]
  \centering
  \includegraphics[width=0.85\linewidth]{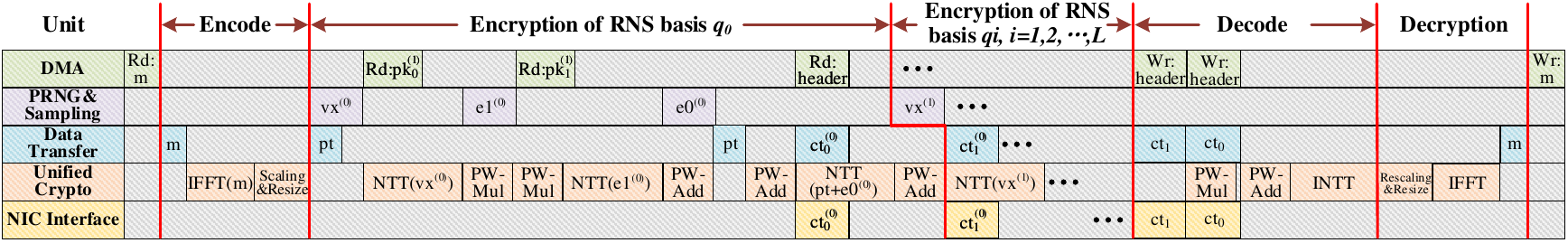}
  \caption{End-to-end PPOC workflow of edge-side RNS-CKKS on the DNA-HHE architecture.}
  \label{fig:Workflow_CKKS}
\end{figure*}
\subsubsection{Workflow of edge-side RNS-CKKS}
Fig. \ref{fig:Workflow_CKKS} demonstrates the end-to-end PPOC workflow of edge-side RNS-CKKS on the proposed architecture.
Benefiting from the conflict management of Task Manager, the independent tasks can be executed in an out-of-order manner to reduce overall latency.
The final ciphertexts, $ct^{i}_{0}$ and $ct^{i}_{1}$, are stored in RAM0 and RAM1 of Key\_BUF0, respectively, replacing the original $pk^{i}_{0}$ and $pk^{i}_{1}$.
RAM0/1 is specifically designed with dual backups to accommodate data from two RNS domains, thus allowing independent tasks in both domains to execute in parallel.
For network packet encapsulation and transmission, firstly we utilize the DAU to read packet headers to the NIC\_BUF4, then the DTU moves the ciphertexts segments from Key\_BUF0 to the position in the NIC\_BUF4 following the packet header, and finally, the NIU controls the packet transfer directly with NIC;
the packet header size and ciphertext segment length can be dynamically set through the configuration instructions to the Config Unit.
In this way, the overhead of ciphertext encapsulation and transmission is significantly reduced, thereby improving the speed of edge-side end-to-end PPOC.

\subsubsection{Workflow of Rubato}
Rubato SE involves diverse parameter configurations, including different block sizes and computation rounds.
Therefore, instead of employing customized execution flow running in cascading units like the PASTA cipher implementation \cite{cryptoeprint:2024/1919}, this work performs Rubato cipher through flexible instruction programming based on fine-grained task partitioning, thereby supporting various parameter settings under a unified architecture.

\section{Evaluation}
\subsection{Experimental Setup}
The proposed DNA-HHE architecture is implemented using Chisel HDL \cite{bachrach2012chisel}, then integrated into the Chipyard SoC \cite{9099108} as an MMIO peripheral via TileLink interconnect protocol \cite{cook2017diplomatic}.
We employ the Rocket core \cite{asanovic2016rocket} as CPU, and IceNet \cite{icenet_chipyard} is utilized as the network interface controller that DNA-HHE is directly coupled to.
Network loopback tests are conducted to ensure correctness against the software references, the FullRNS-HEAAN \cite{CheonHKKS18} implementation of RNS-CKKS, and Rubato \cite{HaKLLS22}.
DNA-HHE is synthesized under TSMC 28nm HPC+ technology using Synopsys Design Compiler and Memory Compiler, and is also instantiated on the Xilinx Kintex-7 KC705 Platform with Vivado.

\subsection{Implementation Results and Comparisons}
The results and comparisons are structured in three parts.
Initially, the performance of DNA-HHE on edge-side RNS-CKKS operations is presented and compared with related state-of-the-art (SOTA) works.
Next, we demonstrate the performance of DNA-HHE on Rubato SE and compare it with relevant studies.
Finally, we showcase the latency benefits of near-network computing in end-to-end PPOC applications.

\subsubsection{Edge-side RNS-CKKS}
\begin{table*}[htb]
\centering
\caption{Performance and comparison with related work of edge-side RNS-CKKS}
\label{tab:res_rns_ckks}
\resizebox{\textwidth}{!}{
\begin{threeparttable}
\begin{tabular}{|c|c|c|c|c|ccccc|ccc|c|}
\hline
\multirow{2}{*}{\textbf{Work}} & \multirow{2}{*}{\textbf{Platform}} & \textbf{Supported}          & \multirow{2}{*}{$\bm{N}$} & \multirow{2}{*}{$\bm{\mathrm{log_{2}}Q}$} & \multicolumn{5}{c|}{\textbf{Area}}                                                                                              & \multicolumn{3}{c|}{\textbf{Time (ms)}}                                                   & \textbf{Area\,\tnote{c}}\,$\bm{\times}$\textbf{Time} \\ \cline{6-13}
                      &                           & \textbf{Mode}               &                    &                       & \multicolumn{1}{c|}{\textbf{LUT}}   & \multicolumn{1}{c|}{\textbf{FF}}    & \multicolumn{1}{c|}{\textbf{DSP}} & \multicolumn{1}{c|}{\textbf{BRAM}} & \textbf{ENS\tnote{a}}   & \multicolumn{1}{c|}{\textbf{m-to-ct\tnote{b}}} & \multicolumn{1}{c|}{\textbf{ct-to-m\tnote{b}}} & \textbf{Total}  & \textbf{ratio}    \\ 
\hline
\hline
SEAL \cite{sealcrypto}              &  Intel i5-8250U @2.4GHz         & RNS-CKKS           & 8192               & 3×54                  & \multicolumn{1}{c|}{--} & \multicolumn{1}{c|}{--} & \multicolumn{1}{c|}{--}  & \multicolumn{1}{c|}{--}   & -- & \multicolumn{1}{c|}{4.91}           & \multicolumn{1}{c|}{0.73}           & 5.64 & --     \\ 
\hline
\hline
RISE\tnote{d} \cite{AzadYAPTJ23} & ASIC 12nm @ 1000MHz & RNS-CKKS & 8192 & 180 & \multicolumn{4}{c|}{113705 $\mu\mathrm{m}^{2}$} & -- & \multicolumn{1}{c|}{20.00} & \multicolumn{1}{c|}{19.00} & 39.00 & 7.38 \\ \hline
\textbf{DNA-HHE (our)} & ASIC 28nm @ 500MHz & RNS-CKKS+Rubato SE & 8192 & 3×54 & \multicolumn{4}{c|}{721480\tnote{e} $\mu\mathrm{m}^{2}$} & -- & \multicolumn{1}{c|}{0.72} & \multicolumn{1}{c|}{0.12} & 0.83 & 1.00 \\ 
\hline
\hline
Aloha-HE \cite{KriegerHMR24}              & Kintex-7 @ 200MHz         & RNS-CKKS           & 8192               & 3×54                  & \multicolumn{1}{c|}{17680} & \multicolumn{1}{c|}{14431} & \multicolumn{1}{c|}{56}  & \multicolumn{1}{c|}{97}   & 36248 & \multicolumn{1}{c|}{2.14}           & \multicolumn{1}{c|}{1.14}           & 3.28 & 1.81     \\ 
\hline
CAEA \cite{WangYHZMSL25}                 & Kintex-7 @ 200MHz         & RNS-CKKS           & 8192               & 3×54                  & \multicolumn{1}{c|}{10804} & \multicolumn{1}{c|}{11688} & \multicolumn{1}{c|}{60}  & \multicolumn{1}{c|}{84}   & 31009 & \multicolumn{1}{c|}{1.78}           & \multicolumn{1}{c|}{0.42}           & 2.20  & 1.04     \\ 
\hline
\textbf{DNA-HHE (our)}         & Kintex-7 @ 200MHz         & RNS-CKKS+Rubato SE & 8192               & 3×54                  & \multicolumn{1}{c|}{18686} & \multicolumn{1}{c|}{10444} & \multicolumn{1}{c|}{36}  & \multicolumn{1}{c|}{91.5} & 31428 & \multicolumn{1}{c|}{1.79}           & \multicolumn{1}{c|}{0.29}           & 2.08 & 1.00        \\ \hline
\end{tabular}
\begin{tablenotes}
    \item[a] The equivalent number of slices (ENS) is defined according to the formula in \cite{10634219}: $\mathrm{ENS} = \mathrm{DSP}\times 100+\mathrm{BRAM}\times 196+\frac{\mathrm{LUT}}{4}+\frac{\mathrm{FF}}{2}$.
    \item[b] m-to-ct and ct-to-m respectively stand for the message-to-ciphertext process including encode and encryption, and the ciphertext-to-message process including decryption and decode.
    \item[c] For FPGA implementations, the area is measured using the ENS metric; for ASIC implementations, the normalized area metric is employed.
    \item[d] This work presents its implementation results exclusively in diagrams, without offering any exact values.
    \item[e] The area consumption is normalized from 28nm to 14nm process according to the area scaling factor presented in \cite{9401196}.
\end{tablenotes}
\end{threeparttable}
}
\end{table*}
Table \ref{tab:res_rns_ckks} demonstrates the performance and comparison with related works of edge-side RNS-CKKS.
We use the equivalent number of slices (ENS) metric to represent the overall area overhead for FPGA implementations and use the ratio of Area$\times$Total Time as a measure of area efficiency.
Compared to the software benchmark running on high-end CPU, our work achieves more than 2.53$\times$ improvement in latency in both the m-to-ct and ct-to-m process.
Compared to RISE, our work achieves a $46.8\times$ speedup in total latency and increases the area efficiency by a factor of $7.38\times$.
RISE employs minimal on-chip memory, which significantly reduces the area overhead but results in fewer data reuse opportunities and higher data transfer costs, thereby increasing its latency.
Compared to Aloha-HE, our work obtains 1.19$\times$, 3.95$\times$, and 1.57$\times$ improvement in latency, respectively, in m-to-ct, ct-to-m, and the overall process.
The performance gains benefit from the out-of-order task execution and two-BFU-parallel FFT/IFFT in this work, while Aloha-HE lacks the ability of task scheduling and only employs one BFU for FFT/IFFT in complex field.
Additionally, our work consumes 1.15$\times$ less ENS than Aloha-HE and achieves 1.81$\times$ better area efficiency.
Compared to CAEA, the m-to-ct latency is increased by 0.69\%, while the ct-to-m latency is decreased by 31.4\%;
overall, our total latency is improved by a factor of 1.06$\times$.
CAEA reduces the number of NTT in the encryption by computing the NTT of the public keys instead of the sampling errors, which adds one additional NTT for the secret key during the decryption in return.
Besides, by omitting the PRNG and sampling units on the accelerator, CAEA lowers area overhead, particularly the LUT usage;
however, this strategy leads to dependence on DMA transfers for obtaining sampling errors, instead of the faster local generation, thus increasing the encryption latency.
Compared to CAEA, our work achieves 1.04$\times$ better area efficiency, benefiting from the proposed DSP-efficient modular reduction, which reduces 40\% DSP usage.
It can be seen that compared to the SOTA single-mode RNS-CKKS accelerators, our work achieves better computation and area efficiency, while additionally supporting the Rubato SE cipher and dual-mode HHE computation.

\subsubsection{Rubato SE}
\begin{table}[htb]
\centering
\caption{Performance and comparison with software of Rubato}
\label{tab:res0_rubato}
\resizebox{\columnwidth}{!}{%
\begin{tabular}{|c|c|ccc|}
\hline
\multirow{2}{*}{\textbf{Work}} & \multirow{2}{*}{\textbf{Platform}} & \multicolumn{3}{c|}{\textbf{Cycle and Speedup}}                                            \\ \cline{3-5} 
                      &                           & \multicolumn{1}{c|}{\textbf{Par-128S}}    & \multicolumn{1}{c|}{\textbf{Par-128M}}    & \textbf{Par-128L}    \\ 
\hline
\hline
Rubato \cite{HaKLLS22}               & AMD Ryzen 7 2700X         & \multicolumn{1}{c|}{10446}       & \multicolumn{1}{c|}{14292}       & 16920       \\ \hline
\hline
\textbf{DNA-HHE (our)}         & Kintex-7 FPGA             & \multicolumn{1}{c|}{1235 (8.5×)} & \multicolumn{1}{c|}{2087 (6.8×)} & 3036 (5.6×) \\ \hline
\end{tabular}
}
\end{table}

\begin{table}[htb]
\centering
\caption{Comparison with related work of symmetric encryption in HHE}
\label{tab:res1_rubato}
\resizebox{\columnwidth}{!}{%
\begin{tabular}{|c|c|c|c|c|c|c|}
\hline
\multirow{2}{*}{\textbf{Work}}           & \multirow{2}{*}{\textbf{Platform}} & \textbf{Supported} & \textbf{Block}  & \textbf{Area}                            & \multirow{2}{*}{\textbf{Cycle}} & $\bm{\mathrm{A}\times\mathrm{C}/n}$ \textbf{ratio}     \\ \cline{5-5} \cline{7-7} 
                                &                           & \textbf{Mode}      & \textbf{Size} $\bm{n}$ & \textbf{LUT/FF/DSP}                      &                        & \textbf{LUT/FF/DSP}      \\
\hline
\hline
\multirow{2}{*}{Aikata \cite{cryptoeprint:2024/1919}} & Artix-7                   & PASTA-4   & 32     & 23.7k/11.1k/64                  & 1591                   & 1.33/1.12/1.86  \\ \cline{3-7} 
                                & @75MHz                    & PASTA-3   & 64     & 65.5k/36.3k/256                 & 4955                   & 5.71/5.67/11.61 \\ 
\hline
\hline
\multirow{3}{*}{\textbf{DNA-HHE (our)}}  & \multirow{3}{*}{\begin{tabular}[c]{@{}c@{}}Kintex-7 \\      @200MHz\end{tabular}} & \multirow{3}{*}{\begin{tabular}[c]{@{}c@{}}RNS-CKKS\\      +Rubato\end{tabular}} & 16     & \multirow{3}{*}{18.7k/10.4k/36} & 1235                   & 1.63/1.63/1.63  \\ \cline{4-4} \cline{6-7} 
                                &                                                                                   &                                                                                  & 36     &                                 & 2087                   & 1.22/1.22/1.22  \\ \cline{4-4} \cline{6-7} 
                                &                                                                                   &                                                                                  & 64     &                                 & 3036                   & 1.00/1.00/1.00  \\ \hline
\end{tabular}
}
\end{table}
Table \ref{tab:res0_rubato} shows the performance of Rubato SE and comparison with the software baseline in high-end CPU, and our work obtains a more than 5.6$\times$ speedup in cycles across all 128-bit security settings.
As our work is the \textbf{\textit{first}} hardware implementation of Rubato, in terms of hardware-based counterparts, we select the hardware accelerator \cite{cryptoeprint:2024/1919} for comparison, which implements the PASTA SE \cite{Dobraunig0HRSW23} tailored for the HHE variant of BFV/BGV.
The comparison results are shown in Table \ref{tab:res1_rubato} with the same 128-bit security level.
Considering the differences in the block size $n$, we utilize the metric $\frac{\mathrm{Area\times Cycle}}{\mathrm{block\;size} \;n}$ to measure the area efficiency.
Under the same block size of $n$ = 64, our work achieves a 1.63$\times$ reduction in latency compared to \cite{cryptoeprint:2024/1919}.
The Par-128L setting with $n=64$ in this work achieves the highest area efficiency, with improvements of over 1.33$\times$ and 1.86$\times$ in LUT and DSP utilization compared to \cite{cryptoeprint:2024/1919}, respectively,
while our work additionally supports edge-side RNS-CKKS operations.

\subsubsection{Near-network Acceleration}
The edge-side end-to-end PPOC applications consist of two processes, the message-to-ciphertext-to-cloud  (m-to-ct-to-cloud) process, and the inverse cloud-to-ct-to-m process, the latency of which are modeled by summing the computation latency of m-to-ct/ct-to-m in user mode and the ciphertext transmission latency in kernel mode.
In the approach of standalone accelerators, ciphertext transmission latency is obtained by summing the latency of ciphertexts Store from the accelerator to memory and the latency of ciphertexts Load from the memory to NIC, via the TileLink interconnect bus.
In our near-network approach, this ciphertext transmission latency can be replaced by the non-overlapping delay of network protocol packaging and transmission, since the accelerator is directly coupled with the NIC and can handle packet en/de-capsulation and sending/receiving itself.
The m-to-ct/ct-to-m delays from this work are used in the total latency for both approaches to maintain consistency in the computational process, thus highlighting the effect of the ciphertext transmission approaches.

\begin{table}[t!]
\centering
\caption{Ciphertext Transfer Overhead via TileLink interconnect bus}
\label{tab:data_transfer}
\resizebox{\columnwidth}{!}{%
\begin{tabular}{|c|c|cc|cc|cc|}
\hline
\textbf{TileLink} & \textbf{TileLink} & \multicolumn{2}{c|}{\textbf{Cycle of}} & \multicolumn{2}{c|}{\textbf{Overall Cycle}} & \multicolumn{2}{c|}{\textbf{Percentage of  Data Transfer}} \\ \cline{5-8} 
\textbf{Channel} & \textbf{Bus} & \multicolumn{2}{c|}{$\bm{N=8192}$ \textbf{words}} & \multicolumn{1}{c|}{\textbf{m-to-ct-}} & \textbf{cloud-to-} & \multicolumn{1}{c|}{\textbf{m-to-ct-}} & \textbf{cloud-to-} \\ \cline{3-4}
\textbf{Number} & \textbf{Width} & \multicolumn{1}{c|}{\textbf{Load}} & \textbf{Store} & \multicolumn{1}{c|}{\textbf{to-cloud}} & \textbf{ct-to-m} & \multicolumn{1}{c|}{\textbf{to-cloud}} & \textbf{ct-to-m} \\
\hline
\hline
\multirow{3}{*}{1} & 64 & \multicolumn{1}{c|}{16924} & 12520 & \multicolumn{1}{c|}{533.5k} & 116.5k & \multicolumn{1}{c|}{33.11\%} & 50.55\% \\ \cline{2-8} 
 & 128 & \multicolumn{1}{c|}{12807} & 12338 & \multicolumn{1}{c|}{505.7k} & 107.8k & \multicolumn{1}{c|}{29.83\%} & 46.65\% \\ \cline{2-8} 
 & 256 & \multicolumn{1}{c|}{10767} & 12343 & \multicolumn{1}{c|}{492.5k} & 103.7k & \multicolumn{1}{c|}{28.16\%} & 44.56\% \\
 \hline
 \hline
\multirow{3}{*}{2} & 64 & \multicolumn{1}{c|}{9882} & 8288 & \multicolumn{1}{c|}{461.9k} & 91.8k & \multicolumn{1}{c|}{23.60\%} & 39.58\% \\ \cline{2-8} 
 & 128 & \multicolumn{1}{c|}{6623} & 6300 & \multicolumn{1}{c|}{428.8k} & 80.3k & \multicolumn{1}{c|}{18.08\%} & 32.17\% \\ \cline{2-8} 
 & 256 & \multicolumn{1}{c|}{5551} & 6187 & \multicolumn{1}{c|}{421.1k} & 77.9k & \multicolumn{1}{c|}{16.72\%} & 30.13\% \\ \hline
\end{tabular}%
}
\end{table}

\begin{figure}[hbt]
  \centering
  \includegraphics[width=0.9\linewidth]{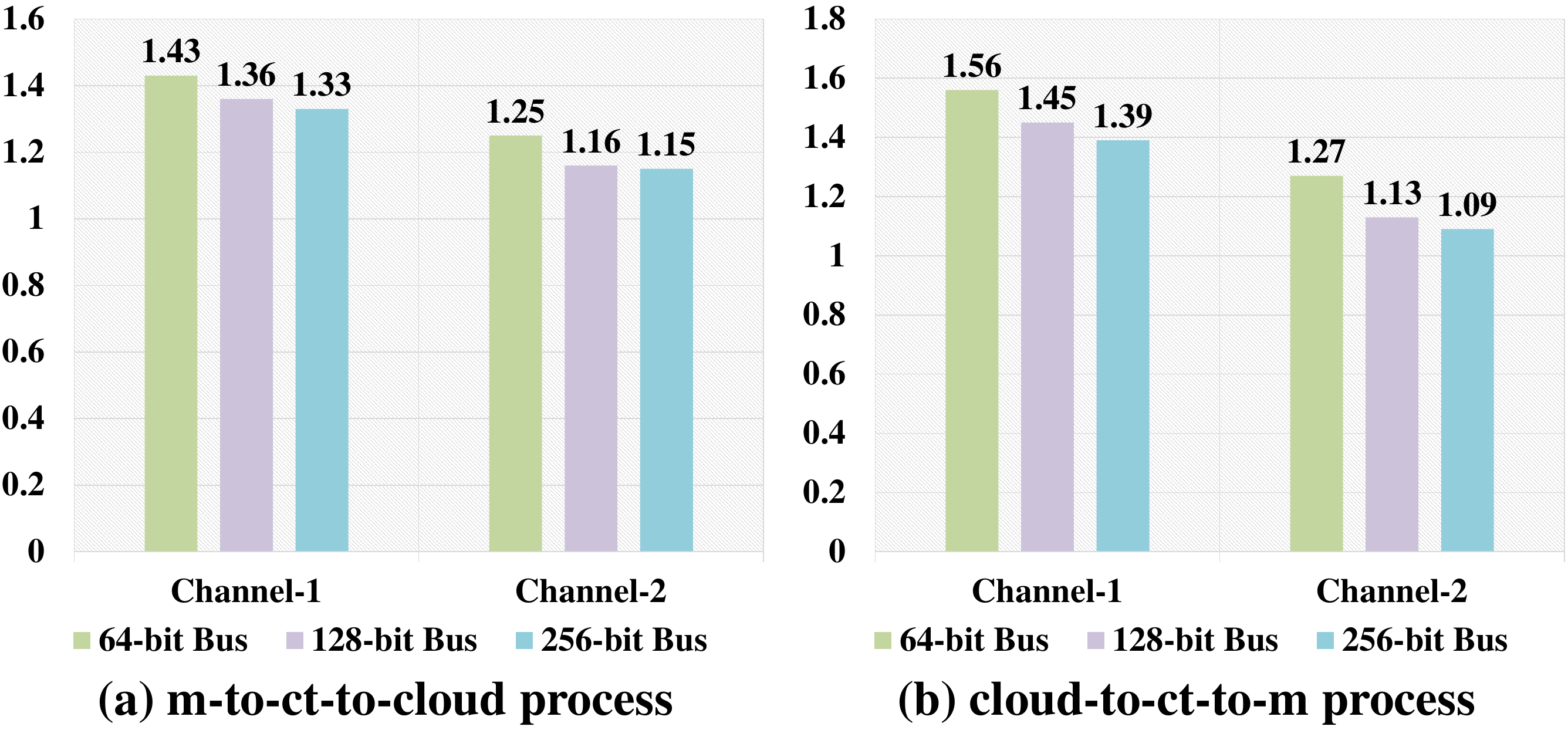}
  \caption{Speedup of the near-network approach compared to the standalone-accelerator approach.}
  \label{fig:near_net_speedup}
\end{figure}

Table \ref{tab:data_transfer} illustrates the ciphertext transfer overhead in the standalone-accelerator approach via TileLink interconnect bus with parameter settings of $N=8192$ and $\mathrm{log_{2}}Q=3\times 54$.
Experiments are conducted on multiple TileLink bus configurations on the edge side, including bus widths ranging from 64 to 256 bits and two options of channel number; having more than 2 channels does not lead to any improvements in DMA speed on the tested SoC.
The overall cycle of the m-to-ct-to-cloud process is simulated by the addition of the m-to-ct latency in this work and the ciphertext transfer latency in $3\times$ RNS domains.
It can be seen that ciphertext transmission constitutes a notable portion of the edge-side end-to-end PPOC latency, accounting for 16.72\% to 33.11\% in the m-to-ct-to-cloud process and 30.13\% to 50.55\% in the cloud-to-ct-to-m process.
Besides, to reduce ciphertext transmission latency, increasing the number of channels from 1 to 2 is more effective than doubling the bus width.
Furthermore, we present the speedup of the near-network approach compared to the standalone-accelerator approach in edge-side end-to-end PPOC in Fig. \ref{fig:near_net_speedup}.
The near-network approach achieves 1.15$\times$ to 1.43$\times$ acceleration in the m-to-ct-to-cloud process, and 1.09$\times$ to 1.56$\times$ speedup in the cloud-to-ct-to-m process, under different TileLink bus configurations.

\vspace{-20pt}
\subsection{Discussion of RNS-CKKS vs. Rubato}
\begin{figure}[hbt]
  \centering
  \includegraphics[width=0.9\linewidth]{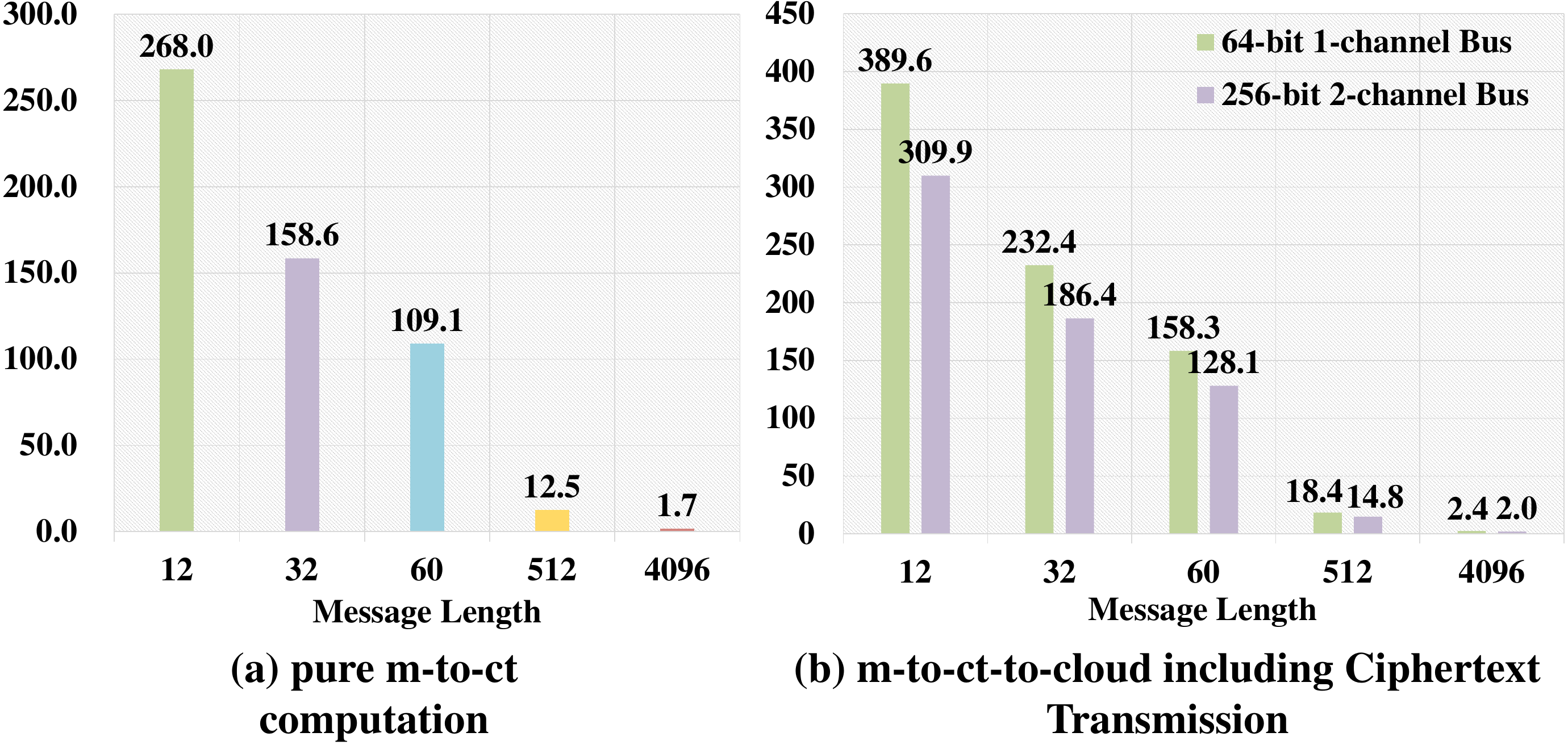}
  \caption{Speedup of Rubato SE compared to the RNS-CKKS.}
  \label{fig:rubatoVsFHE}
\end{figure}
In Fig. \ref{fig:rubatoVsFHE}(a) and (b), we respectively present the speedup in latency of Rubato SE compared to the RNS-CKKS in pure m-to-ct computation and end-to-end m-to-ct-to-cloud process including the ciphertext transmission.
To ensure a thorough discussion, we select three levels of message lengths $L_{m}$:
short-length messages matching the output size of Rubato ($L_{m}=12,32,60$), medium-length message with $L_{m}=512$, and the maximum message length $L_{m}=N/2=4096$ that can be batch-processed with RNS-CKKS under the $N$=8192 setting.
Additionally, we choose the slowest and fastest TileLink bus types shown in Fig. \ref{fig:near_net_speedup} to determine the upper and lower bounds of the speedup ratio.
To ensure a fair comparison, the latency for Rubato SE and RNS-CKKS are both collected from the proposed DNA-HHE accelerator, with both algorithms using the same level of two-way computational parallelism.

As shown in Fig. \ref{fig:rubatoVsFHE}, Rubato outperforms RNS-CKKS across all levels of message lengths.
For short messages, Rubato demonstrates a latency advantage of more than 100$\times$.
However, as the message length increases, the speedup ratio decreases to several tens for medium-length messages, and eventually reduces to a few for long messages; this decline is attributed to the batch processing capability of RNS-CKKS.
Besides, compared to the pure m-to-ct computation, Rubato exhibits a higher acceleration ratio in the end-to-end m-to-ct-to-cloud process that includes ciphertext transmission overhead, because Rubato SE avoids the ciphertext expansion problem in RNS-CKKS, thereby reducing the transmission costs.
From the discussion above, we can summarize the most suitable scenarios respectively for Rubato and RNS-CKKS, thereby guiding the algorithm switching and practical deployment of PPOC on the edge side.
Rubato is better suited for the following scenarios:
\begin{itemize}
\item 
When the batch size of messages is relatively small, especially for short ones with a length of less than 100, such as the personal-level PPOC for medical diagnosis \cite{7820321} and financial transactions \cite{haryaman2024secure}.
\item 
When edge-side resources are limited, particularly in cases where system interconnection or networking has low capability or congestion.
\item 
When cloud computing resources are sufficient and the overhead of transciphering is acceptable.
\end{itemize}
On the other hand, RNS-CKKS is more suitable in these scenarios:
\begin{itemize}
\item 
When the batch size of messages is relatively large, especially for long ones with a length exceeding 1000, such as the secure big data analysis \cite{alabdulatif2020towards} and PPOC for long-token inference of large language models \cite{ruoyan2025practical}.
\item 
When the bandwidth of system interconnection and networking is relatively sufficient.
\item 
When cloud computing resources are insufficient.
\end{itemize}

\section{Conclusion}
In this paper, we present DNA-HHE, the \textbf{\textit{first}} dual-mode HHE accelerator for edge-side RNS-CKKS and Rubato SE cipher.
For resource-constrained edge devices, DNA-HHE is designed to achieve a balanced trade-off between functionality, performance, and area cost  in a unified architecture by adopting a compact multi-field-adaptive BFU, efficient operation scheduling, and out-of-order task processing.
To reduce the ciphertext transmission overhead, DNA-HHE is further equipped with near-network protocol packaging and transmission ability.
It is worth mentioning that the design methodology of DNA-HHE is not limited to the specific parameter configuration but can be extended to various application setups.
We conduct experimental evaluations of DNA-HHE on ASIC and FPGA platforms, which shows that it outperforms existing works while offering dual-mode functionality.
We assess the ciphertext transmission overhead in the edge-side end-to-end PPOC processes and demonstrate the latency benefits of our near-network computing approach.
Moreover, we further analyze the performance comparison between RNS-CKKS and Rubato and identify their respective suitable scenarios.
In future work, we will focus on system-level optimizations for the transciphering overhead involved in HHE schemes on the cloud side.

\ifCLASSOPTIONcaptionsoff
  \newpage
\fi



%


  \bibliographystyle{IEEEtran}
  \bibliography{IEEEabrv,mylib}


%




\end{document}